# A simplified method for full-wave simulation of metamaterials: utilizing near-field decoupling technology

Junming Zhang, Weijia Luo, Yongzheng Wen, Jingbo Sun, Ji Zhou[*]

State Key Laboratory of New Ceramics and Fine Processing, School of Materials Science and Engineering, Tsinghua University, Beijing 100084, China

E-mail: zhouji@tsinghua.edu.cn

Abstract

Simulating the electromagnetic properties of large-scale, complex metamaterial structures demands significant time and memory resources. If these large-scale structures can be divided into smaller, simpler components, the overall cost of studying all the smaller structures could be much lower than directly simulating the entire structure. Unfortunately, decoupling complex structures has been challenging due to the unclear mechanisms of near-field coupling in metamaterials. In this paper, we identify that the key to understanding near-field coupling in metamaterials lies in evanescent wave interactions, which can be captured through full-wave simulations. Our findings suggest that by accounting for the influence of evanescent waves, it becomes possible to analytically decouple and then recouple structures, even when the types of metamaterial structures vary. Building on this insight, we successfully decomposed complex structures into multiple groups of simpler components. By studying these simpler components, the electromagnetic properties of the entire structure can be calculated analytically. This decoupling method dramatically reduces the computation time or memory required for research into the electromagnetic properties of metamaterials.

1. Introduction

The rapid development of metamaterials is closely tied to advancements in full-wave simulations [1-5]. Since most metamaterials lack analytical solutions, numerically solving Maxwell's equations through full-wave simulations is the most effective way to study them. This process involves discretizing the metamaterial structure within a solution space, analyzing the field distribution at each node in the grid, and post-processing the results to predict the structure's scattering characteristics. With the support of electromagnetic wave simulation technology, increasingly novel electromagnetic properties are being achieved through intricate structural designs [6-8]. However, as metamaterial structures become more complex—evolving from simple periodic designs to non-periodic ones [9,10], from single-layer to multilayered configurations [11,12], and from ordered to disordered arrangements [13-15]—the scale of numerical simulations expands significantly. This leads to increased computational challenges, including higher memory demands and longer solution times, placing a heavy burden on the simulation process. As metamaterial structures continue to grow in complexity, this trend could impact the efficiency of electromagnetic simulations and push the limits of hardware capabilities, potentially rendering some simulations unsolvable. While more powerful computing resources, improved hardware, and better numerical algorithms can mitigate these challenges to some extent, they cannot fully resolve the fundamental issue of escalating computational demands as simulation scales continue to increase.

From a geometric perspective, complex large-scale structures in metamaterials can be seen as the result of near-field interactions between simpler structures. A practical approach to studying these complex systems is to decompose them into multiple simpler components for individual analysis [16]. Since simpler structures are smaller and involve fewer grid points, they have lower computational complexity, which can significantly reduce overall computational costs. Solving several simple structures is generally more efficient than addressing a single complex structure. However, the effectiveness of this approach relies on whether the scattering characteristics of the simpler structures can accurately predict those of the complex system. This depends on achieving effective near-field coupling and decoupling of the metamaterials—an essential yet challenging task in the field. Current theoretical models, such as effective medium theory [17], transmission line theory [18,19], equivalent circuit methods [20,21], multiple reflection models [22], and coupled mode theory [23-25], each have limitations in addressing this challenge. The unpredictability of near-field interactions often stems from the complex and unknown resonances that occur between different metamaterial components during near-field coupling [26-30]. These resonances are difficult to quantify, presenting a significant obstacle to reliable near-field decoupling and coupling. Similarly, from a mathematical and simulation perspective, no method currently exists that consistently provides reliable near-field decoupling and coupling.

In this paper, we demonstrate that the unique resonances observed in metamaterials are caused by evanescent waves in the near-field and can be effectively analyzed using full-wave simulations. Complex structures can transform propagating waves into evanescent waves. When a traveling wave interacts with metamaterials, it generates evanescent waves in the near-field, which in turn excite other metamaterials within that range. These interactions can be analyzed by numerically solving Maxwell's equations. By studying the transformation between propagating electromagnetic wave modes and evanescent modes induced by metamaterials, we can construct a near-field scattering matrix to capture this behavior. This allows us to predict the scattering properties of complex metamaterials using the near-field scattering matrices of simpler metamaterials. We conducted a comprehensive investigation of three distinct types of metamaterials: disordered metamaterials [31,32], chessboard metamaterials [33], and classic metasurfaces [34]. The study demonstrates that both similar and different types of metamaterials can effectively couple within the near-field, enabling successful physical coupling and decoupling. Based on this decoupling approach, complex structures can be decomposed into simpler components. For example, we designed a disordered metamaterial and compared direct simulations of the entire structure with simulations in which the structure was decoupled into two or four simpler components. The results show that decomposing a large-scale structure into smaller structures can significantly reduce computing resource requirements. As the number of decoupled structures increases, runtime consumption and memory usage are further optimized. This structural decoupling method is compatible with other computational acceleration tools, providing a versatile approach to addressing the challenges of simulating complex metamaterial structures.

2. Theory, modeling and simulation

## 2.1. Far-field scattering matrix

Disordered metamaterials were selected for demonstration due to their complex structure and scattering properties, representing a general case of metamaterials. To simplify the discussion, the disordered metamaterial was placed within a metal rectangular waveguide [35,36]. Far-field coupling and decoupling, which can be achieved using a far-field scattering matrix, are relatively straightforward and will be discussed first.

Figure 1a depicts the schematic diagram for dividing the far-field coupled large-scaled structure into two relatively small-scaled structure. The scattering property of large-scaled structure could be calculated based on the scattering property of two substructures. The scattering matrix $S_{comp}$ of a composite metamaterial resulting from the far-field coupling of different metamaterials can be derived from the scattering matrices $S_A$ and $S_B$ of the individual substructures as following:

First, obtain the scattering matrix for each individual metamaterial $S_A$ and $S_B$. Next, convert the scattering matrix into a transfer matrix $M_A$ and $M_B$. All elements in the transfer matrix M be calculated as follows [35,37]:

$$M_{11} = t - r't'^{-1}r \tag{1}$$

$$M_{12} = r't'^{-1} \tag{2}$$

$$M_{21} = -t'^{-1}r \tag{3}$$

$$M_{12} = t'^{-1} \tag{4}$$

The transfer matrix of the composite metamaterial can be obtained as follows:

$$M_{comp} = M_B * M_A \tag{5}$$

Finally, the scattering of the composite metamaterial could be derived from the transfer matrix as follows:

$$r = -M_{22}^{-1}M_{21} \tag{6}$$

$$t' = M_{22}^{-1} \tag{7}$$

$$t = M_{11} - M_{12}M_{22}^{-1}M_{21} \tag{8}$$

$$r' = M_{12}M_{22}^{-1} \tag{9}$$

For disorder metamaterials, the scattering properties should be describing by matrix form. Figure 1b is the transmission matrix of electromagnetic waves after interaction with disordered metamaterials in a rectangular waveguide under far-field conditions. The electromagnetic wave propagates in the x direction, confined by metal walls on the upper and lower sides in the y direction. Disordered metamaterials are represented by randomly distributed cylinders, visualized as circles in Figure 1a.

In a rectangular waveguide of specific dimensions, any arbitrary field distribution input or output in the far-field can be decomposed into a linear superposition of all propagation modes of the waveguide. The transmission characteristics of the disordered metamaterial can be represented in matrix form, where $t_{m,n}$ denotes the transmission coefficient from mode $n$ in the incident wave to mode $m$ in the transmission wave. Assuming the rectangular waveguide supports $N$ propagation modes, the transmission matrix takes the shape of $N \times N$. Additionally, the other three scattering parameters $r$, $r'$ and $t'$ are also represented in $N \times N$ matrix forms. Consequently, the scattering matrix becomes a 2N × 2N matrix, structured as follows: $S = \begin{bmatrix} r & t' \\ t & r' \end{bmatrix}$.

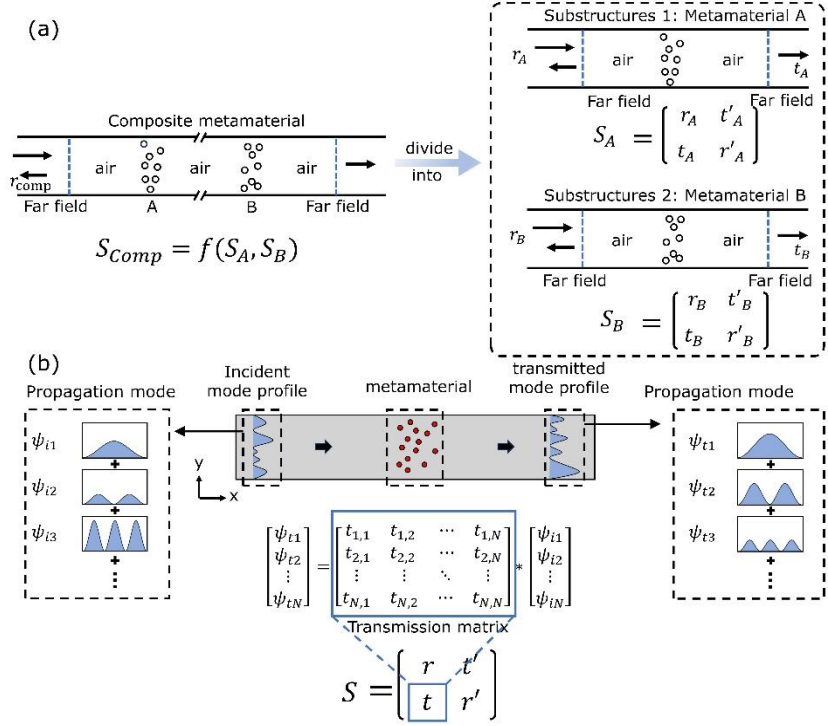

Fig.1. (a) Schematic diagram of complex structure divided into two substructures in far-field scenario; (b) Construction of scattering parameters in matrix form.

2.2 Near-field scattering matrix

For the complex metamaterial shown in figure 2a, it should be divided into four parts: air 1, metamaterial A, metamaterial B, air 2. The coupling for two different metamaterial is the issue to realize the scattering property of complex metamaterial and the issue of coupling for metamaterial and air is to realize the conversion from near to far field. Both of these two issues can be classified to near-field coupling problem.

In the near field, as illustrated in Figure 2b, using a scattering matrix intended for far-field applications will lead to incorrect calculations. This discrepancy arises because when an electromagnetic wave interacts with a metamaterial in the near field, the resulting outgoing wave consists of both propagation and evanescent modes. This is in contrast to the far-field scenario, where the wave can be accurately described as a linear superposition of propagation modes alone.

The presence of evanescent wave modes presents challenges in constructing the scattering matrix due to their theoretically infinite number. However, in practical terms, higher-order evanescent modes decay rapidly. This rapid decay means that once a reference plane is established, the intensity of these higher-order evanescent waves diminishes quickly, rendering their influence negligible in subsequent interactions with the metamaterial. As a result, only a finite number of modes, denoted as $N'$, effectively interact with the metamaterial, including both propagation and evanescent modes. These modes significantly influence interactions in the near field. Therefore, while the theoretical possibility of infinite evanescent modes exists, practical applications focus on this finite set of modes that are crucial for describing the metamaterial's electromagnetic response in the near field [38].

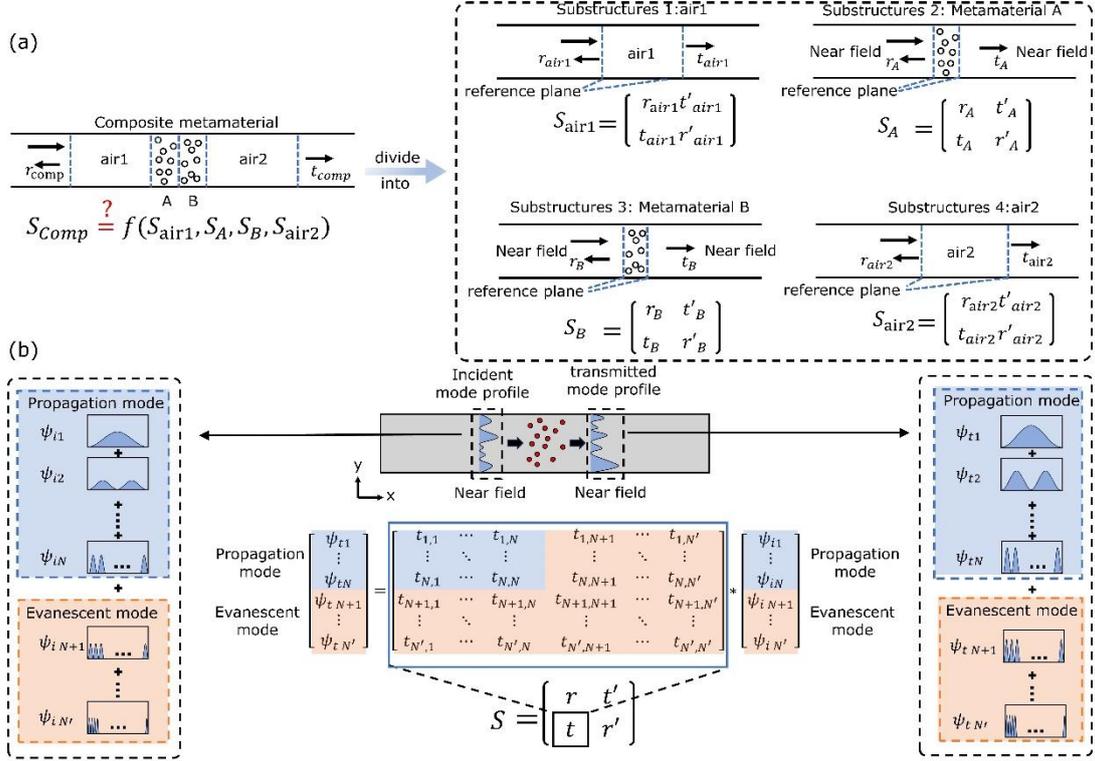

Fig.2. (a) Schematic diagram of complex structure divided into two substructures in near-field scenario; (b) Construction of near-field scattering parameters in matrix form.

2.3. Simulation of the near-field scattering matrix

   To construct the full near-field scattering matrix, it is crucial to calculate all elements that represent the conversion coefficients between propagating modes and evanescent modes. Here, Finite Difference Frequency Domain (FDFD) is selected as demonstrating for its advantages in scenarios with multiple inputs and outputs. FDFD allows for convenient investigation of metamaterial interactions using different source modes, including both propagating and evanescent modes.

   As an example, consider a rectangular waveguide with a cross-section of 100 mm × 8 mm, as illustrated in Figure 3a. At a frequency of 7 GHz, it supports four propagating modes: TE10, TE20, TE30, and TE40. All higher-order modes are evanescent. The spatial distributions of these modes in the simulation space can be obtained by eigenmode calculations using analytical solutions or by eigenmode simulations in simulation software. The $Ez$ components of the first six modes in this rectangular waveguide are shown in Figure S1. Once the source is input into the disordered metamaterial model, the $Ez$ field distribution after interaction with the metamaterial can be obtained through FDFD analysis. From this distribution, the components of different modes in the scattering field can be extracted to obtain the elements of the near-field scattering matrix. In this work, we used FDFD frame from the open-source software MESTI [39] for the simulations.

3. Coupling realizing

3.1 Disorder metamaterial

The disordered metamaterial is created by randomly arranging dielectric cylinders with a radius of 3 mm and permittivity of 2. Two distinct structures, labeled as A and B, are generated and depicted in Figures 3c and 3d, respectively. Each metamaterial consists of 20 cylinders distributed within a 30 mm range. There is a 3 mm air gap on both sides of the structures. At a frequency of 7 GHz, the wavelength in air is approximately 42.86 mm. The 3 mm air gap is much smaller than the wavelength, placing the system in a near-field situation. Through simulation, we obtained the near-field scattering matrices for structures A and B, which are plotted as figure form as shown in figures 3c and 3d.

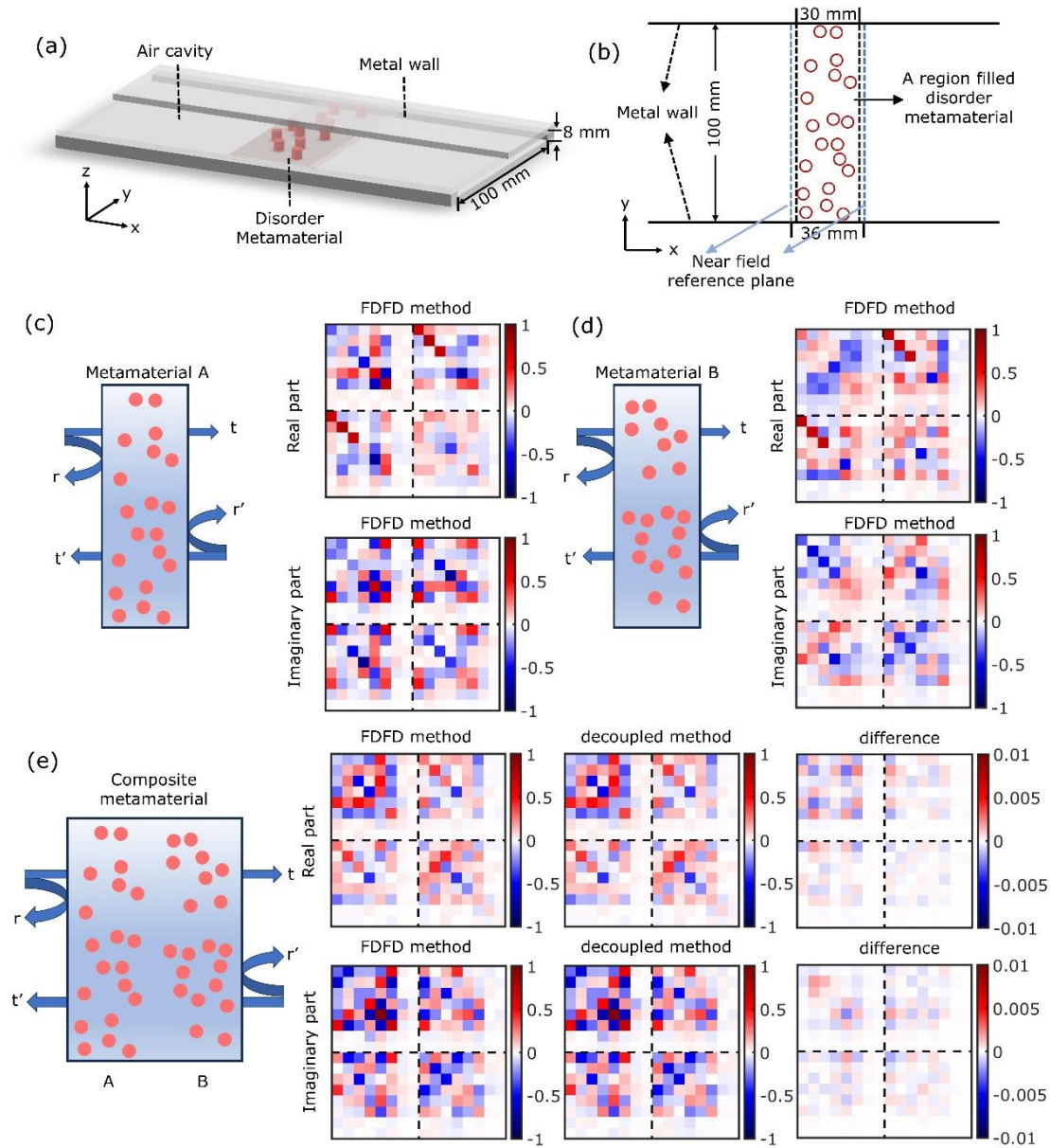

Figure 3. (a) Schematic diagram of disordered metamaterials in a metal rectangular waveguide with a cross-sectional size of 100mm*8mm. (b) is a top view of waveguide; (c) and (d) are the structures of randomly generated disordered metamaterials A and B and their solution area renderings in FDFD, as well as the real and imaginary parts of near-field scattering matrix; (e) the composite metamaterial with its scattering matrix calculated from different method, as well as their difference.

Since the exact number of evanescent modes is hard to predetermine, initially, a large number of modes is set and adjusted based on the resolved scattering matrix. The selected number of modes $N'$ significantly impacts coupling accuracy, detailed discussions of which are provided in the supplementary materials. Typically, a larger initial value is chosen, and the modes that substantially influence the near field are determined after solving to define the dimension of the entire scattering matrix. For the structures shown in Figures 3c and 3d, it has been verified that setting $N'$ to 8 is appropriate, as higher-order modes have minimal impact on these configurations. The near-field scattering matrices obtained from FDFD simulations are depicted in Figures 3c and 3d as 16×16 matrices. By substituting these two field scattering matrices into formulas 1–9, the scattering matrix of the composite structure can be computed. The results align with the simulated outcomes, as illustrated in Figure 3e.

3.2 Chessboard metamaterials

The chessboard metamaterial is also investigated. This type of metamaterial consists of alternating arrangements of isotropic homogeneous materials, as shown in Figure 4a. The length of the metamaterial is set to 15 mm, with its near-field reference set to 3 mm on each side. We designed two different chessboard metamaterials, as shown in Figures 4c and 4d, respectively. For this case, the number of modes is set to 15, resulting in a near-field scattering matrix of size 30 × 30. The near-field coupling structure and its scattering matrix are shown in Figure 4e. The results obtained from the FDFD and near-field coupling calculations match very well.

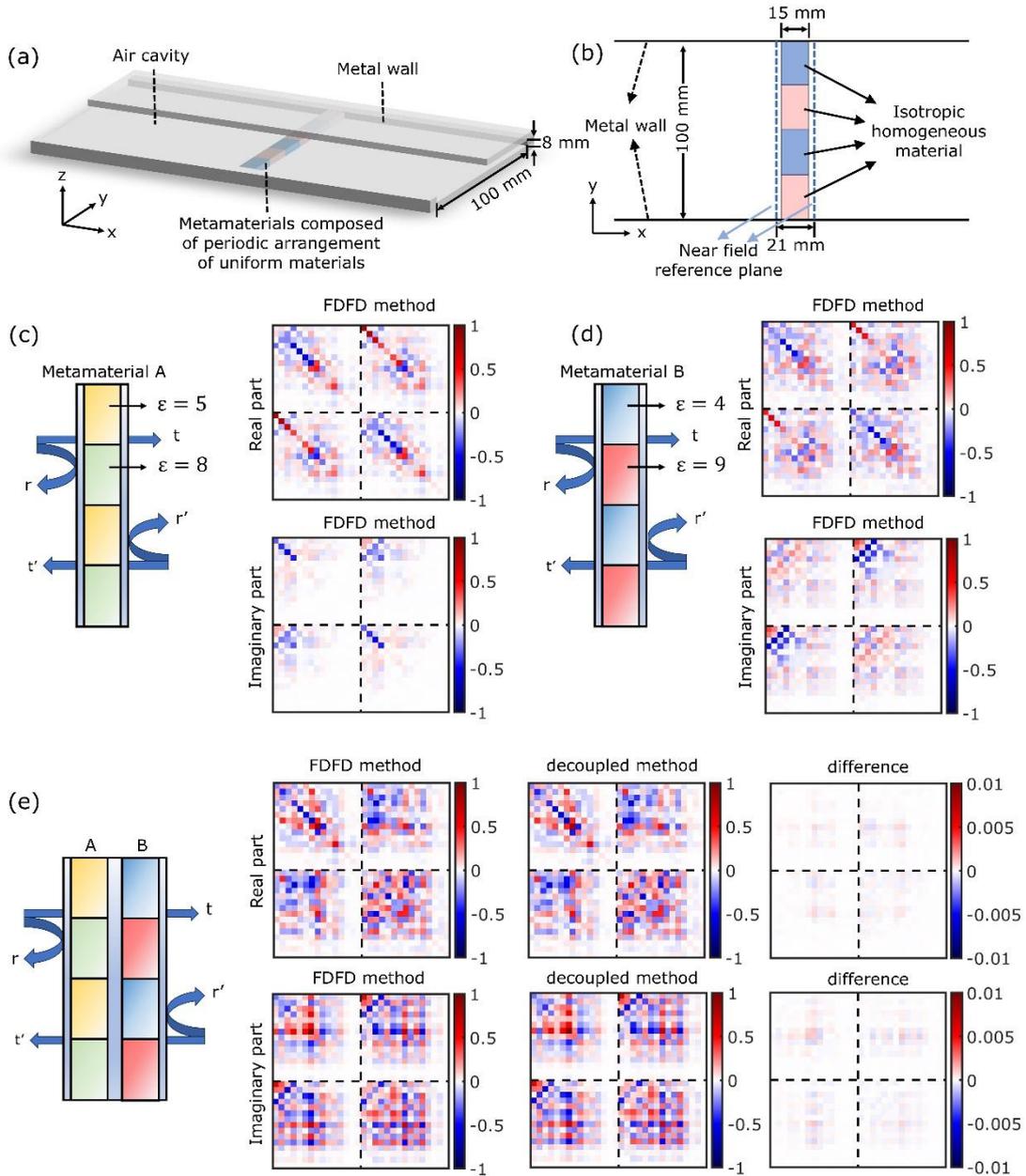

Figure 4. (a) Schematic diagram of the chessboard metamaterial in a metal rectangular waveguide with a cross-sectional size of 100 mm × 8 mm; (b) Schematic diagram of the near-field reference plane for this case; (c) and (d) show the chessboard metamaterials A and B, respectively, along with their solution area renderings in FDFD, as well as the real and imaginary parts of the near-field scattering matrix; (e) The composite metamaterial with its scattering matrix calculated using different methods, as well as the differences between them.

3.3 Metasurfaces

Metasurfaces are a class of metamaterials consisting of periodic arrangements of subwavelength arrays. Here, the units of the metasurface are set as simple structure, as shown in Figure 5a and 5b. The thickness of the metasurface is small compared to the disordered and chessboard metamaterials, so the near-field reference plane is set smaller. In the FDFD simulation, the permittivity of the unit is set to -10,000 to simulate metal, and the permittivity of the substrate is set to 2. By changing the periodicity width, two metasurfaces are built, as shown in Figures 5c and 5d, respectively. For this case, the number of modes is set to 30, resulting in a near-field scattering matrix of size 60 × 60. The near-field coupling results obtained from different methods match very well.

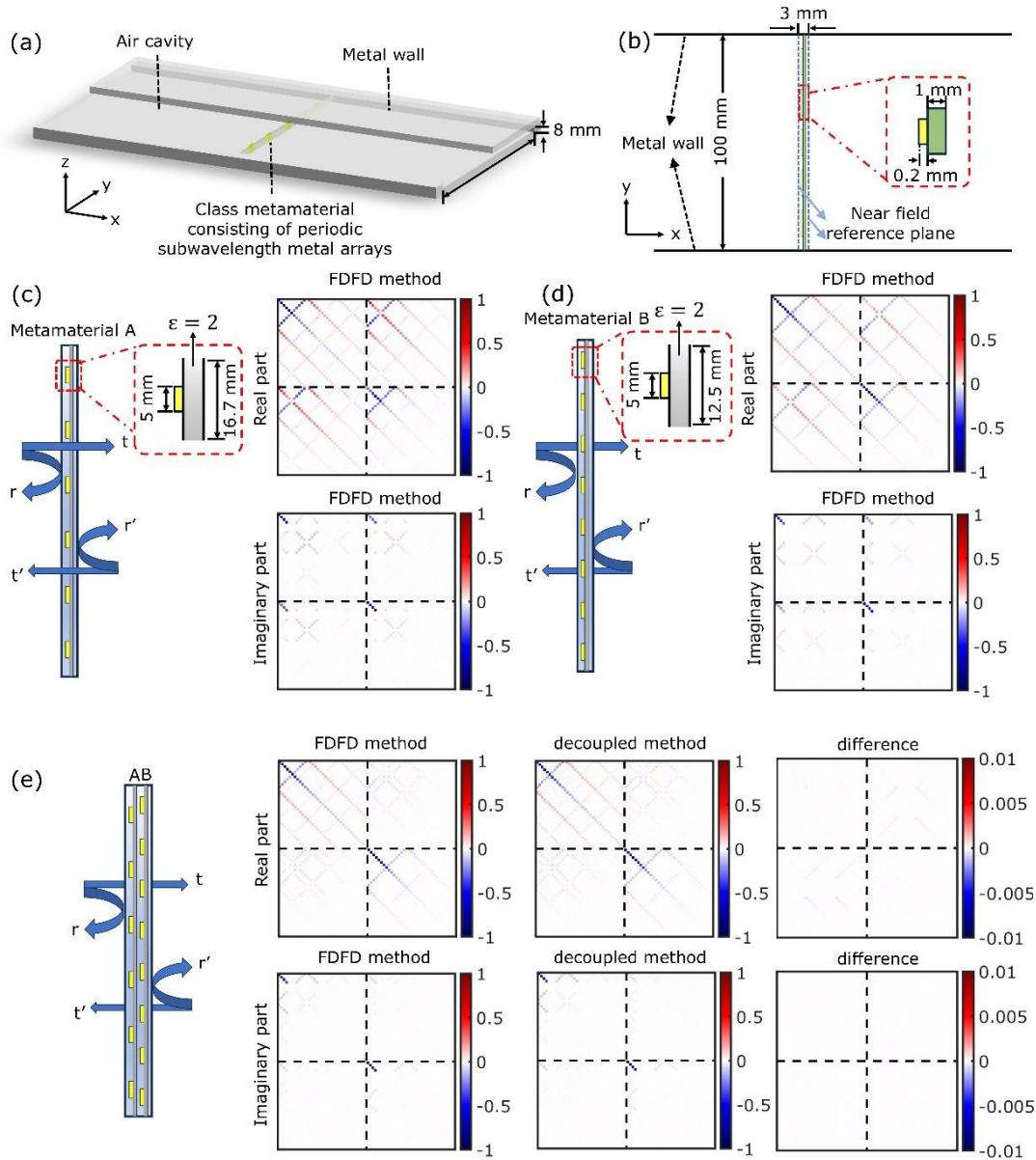

Figure 5. (a) Schematic diagram of the metasurface in a metal rectangular waveguide with a cross-sectional size of 100 mm × 8 mm; (b) Schematic diagram of the near-field reference plane for this case; (c) and (d) show metasurfaces A and B, respectively, along with their solution area renderings in FDFD, as well as the real and imaginary parts of the near-field scattering matrix; (e) The composite metamaterial with its scattering matrix calculated using different methods, as well as the differences between them.

3.4 Coupling for different kind of metamaterials

Near-field coupling between different types of metamaterials can also be achieved. Figure 6a presents a schematic diagram of near-field coupling between various types of metamaterials. The disordered metamaterial, chessboard metamaterial, and metasurface correspond to the structures in Figures 3c, 4c, and 5c, respectively. Figure 6b shows the near-field scattering matrices obtained by FDFD simulations and coupling method, demonstrating good consistency and validating our approach.

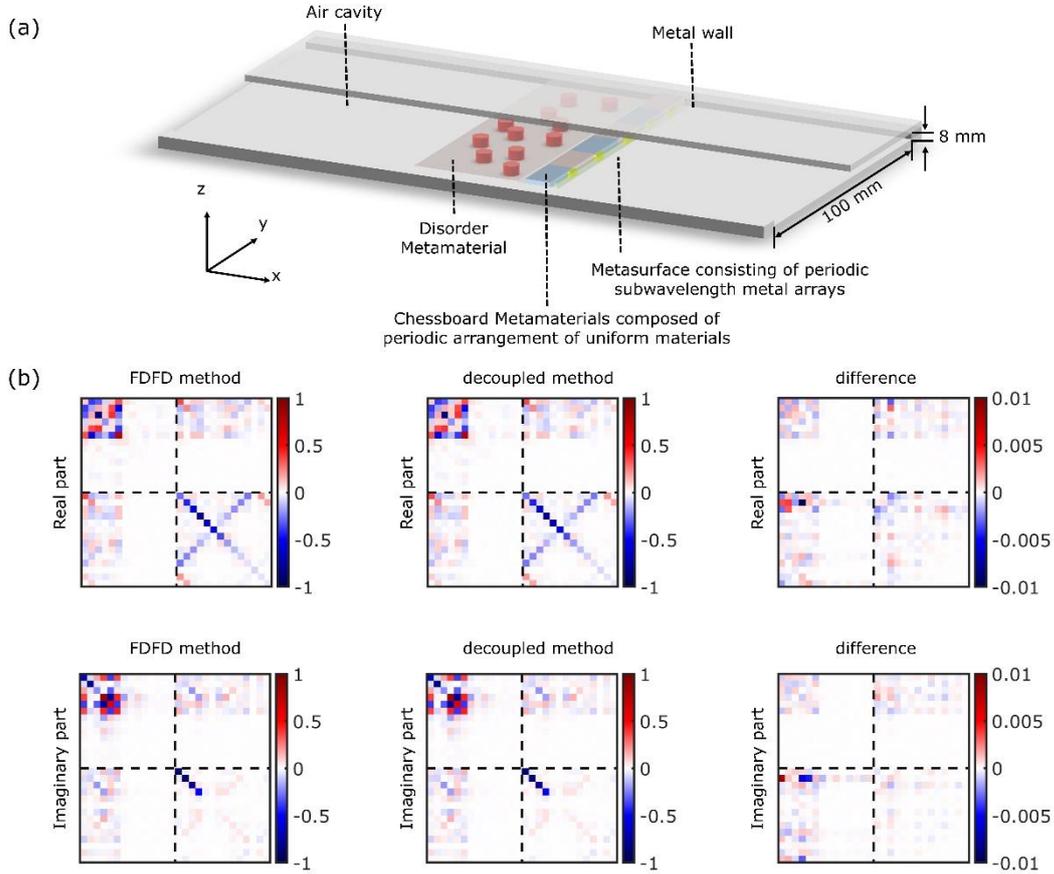

Figure. 6. (a) Schematic diagram of different metamaterials near-field coupled in a metal rectangular waveguide with a cross-sectional size of 100mm*8mm; (c) the near-field scattering matrix of composite metamaterial calculated from different method, as well as their difference.

3.5 Coupling for metamaterials and air

In addition to the coupling between the near fields, our method also workable in the near-field-to-far-field conversion. This process can be performed by coupling the metamaterials with long air gaps on either side. Figure 7a shows the schematic diagram for near-field and far-field reference planes. Figure 7b presents the scattering matrix for an air cavity with a length of 100 mm, where the reflection matrices $r$ and $r'$ are zero matrices, and the transmission matrices $t$ and $t'$ contain elements corresponding only to the four propagation modes, as evanescent waves cannot travel such long distances. By combining air cavities with the disordered metamaterial at the near-field reference plane shown in Figure 3c, the far-field scattering matrix can be obtained, as illustrated in Figure 7c. The results simulated using FDFD are consistent with those calculated from near-field coupling between the metamaterial and long air gaps. Far-field scattering has been widely studied in recent years, and the scattering matrix can be simulated using finite element method. We also studied the structure shown in Figure 7c using COMSOL (based on finite element method) and obtained the same scattering matrix as that coupled with the near-field, further proving the reliability of our method (see detail in supplement material).

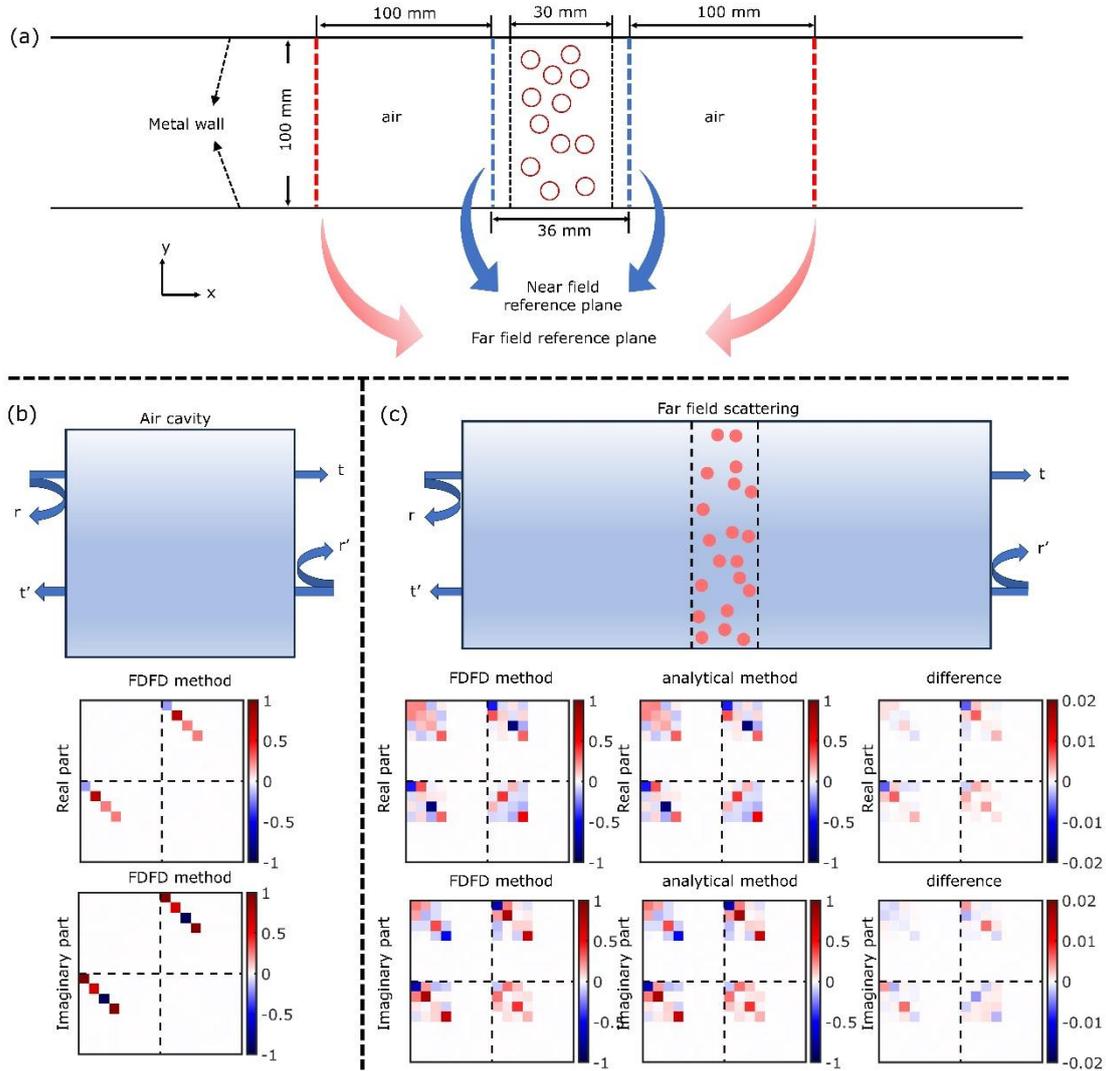

Figure. 7. (a) Schematic diagram of disordered metamaterial with different reference plane. (b) the air cavities in rectangular waveguide and its scattering matrix; (c) the scattering matrix of metamaterial in far-field reference plane calculated from different method, as well as their difference.

3.6 Comparison of simulation efficiency for decoupling method and conventional method

Achieving near-field coupling enables decoupling, allowing large-scale complex structures to be broken down into simpler components. From a full-wave simulation perspective, smaller structures require less memory and lead to faster computations when using the same mesh division method. Therefore, by decoupling large-scale structures, simulation efficiency can be significantly enhanced.

Figure 8a shows a disordered structure. Assuming the grid sizes in the x and y directions are dx and dy, and Nx and Ny represent the number of grids in the x and y directions, respectively, the total number of grids is Nx×Ny. Typically, to determine the scattering characteristics of the structure, the entire structure must be modeled in software and divided into Nx×Ny grids. Based on this grid scale, the field distribution at each grid point is calculated to derive the scattering characteristics.

As the grid size increases, the computation time and memory usage also increase. We meshed the complete structure shown in Figure 8a, resulting in a total of 1,650,327 grid points as the base grid scale, and then adjusting the dx and dy to multiplied the number of grid points. We analyzed the corresponding memory usage and computation time. Figures 8b and 8c compare the calculation time and memory usage as the network size grows multiply.

By implementing decoupling, as shown in Figure 8a, we can divide the metamaterial into multiple substructures and calculate the scattering characteristics of the entire structure by analyzing the scattering behavior of each substructure. Compared to calculating the entire structure at once, this segmentation reduces the overall scale, leading to lower computation time and memory usage. The total time required to calculate all substructures individually is less than that needed to simulate the entire structure, and memory consumption is significantly reduced. For example, when the structure is divided into two equally sized substructures, the total calculation time is approximately 84% of what is required for the complete structure, while memory usage decreases to 46%. When divided into four equally sized substructures, calculation time drops to around 73%, and memory usage is reduced to just 21% of that needed for the complete structure. This optimization remains consistent as the network scale increases, as shown in Figures 8b and 8c. If parallel computation is applied to all substructures, the calculation time can be further reduced. For instance, when the structure is divided into two equally sized substructures, parallel computation reduces memory usage to approximately 93% of that required for the complete structure, with computation time reduced to around 42%. When divided into four substructures, memory usage drops to about 83%, and computation time to just 18% of that required for the complete structure.

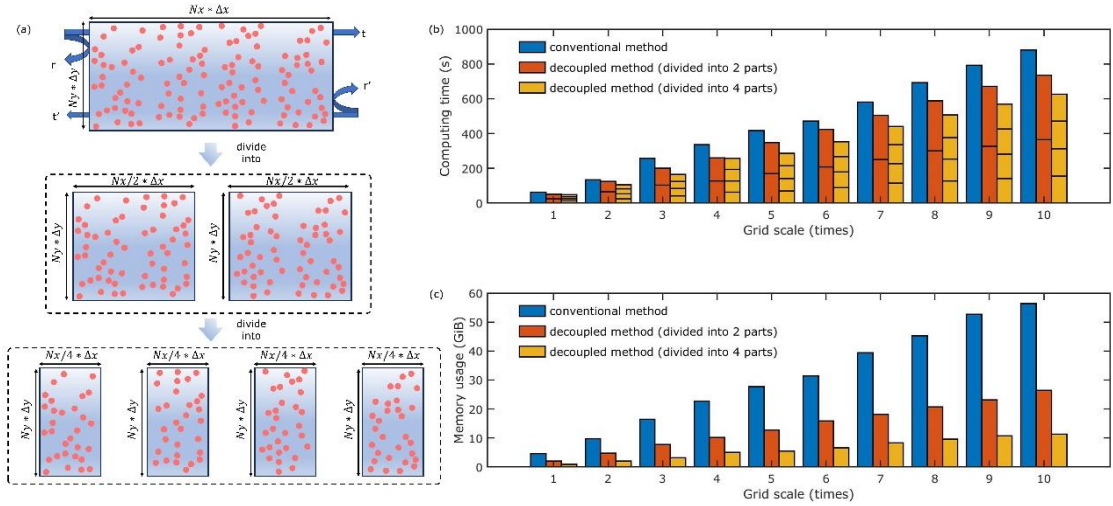

Figure. 8. (a) Schematic diagram of dividing a complex large-scale structure into multiple substructures; (b) Comparison of the simulation time for fully calculating the entire structure and the total time for dividing it into different substructures under different network scales; (c) Comparison of the memory usage for fully calculating the entire structure and the total memory usage for dividing it into different substructures under different network scales.

4. Conclusion

We studied three different types of metamaterials: disordered metamaterials, chessboard metamaterials, and classical metasurfaces. The results demonstrated that near-field decoupling and recoupling can be achieved not only within the same type of metamaterial but also across different types. By coupling air with metamaterials in the near field, a conversion between near-field and far-field scattering characteristics is achieved. Additionally, we compared the time and memory usage of simulating the complete structure versus using the decoupling method. Our findings reveal that, regardless of the scale of mesh division, dividing a complex structure into two equally sized substructures reduces simulation time by an average of 16% and memory usage by 56% when the substructures are run sequentially, compared to directly simulating the complete structure. When parallel operations are used, memory usage is reduced by approximately 7%, and running time is reduced by around 58%. Moreover, dividing the structure into four equally sized substructures can save about 27% of time and 79% of memory when run sequentially, and reduce memory usage by 17% and running time by an average of 82% when using parallel operations.

Our research offers a way to reduce the cost of studying metamaterials in ultra-large-scale complex systems. We believe this method is not only applicable to metamaterials but may also be feasible for other electromagnetic scattering structures. Additionally, this decoupled simulation method is highly beneficial for optimizing metamaterial designs. Since identical parts of the structure no longer need to be included in iterative simulations, the iteration cost can be reduced, leading to faster metamaterial design. The near-field scattering matrix we proposed also enables the correlation of scattering characteristics across different metamaterials. This matrix provides a physical parameter that can be compared and calculated for various metamaterial structures, facilitating communication between different structures. This advancement could potentially contribute to the realization of a "metamaterial genome."

Data availability

The data that support the findings of this study are available from the corresponding authors upon reasonable request.

Acknowledgements

This work was supported by National Natural Science Foundation of China (52332006), the National Key R&D Program of China (2022YFB3806000，2023YFB3811401), the Southwest United Graduate School Research Program (202302AO370008), the China Postdoctoral Science Foundation (Grant No. 2023M742006) and the National Natural Science Foundation of China (grant number 52202370).

Conflict of interest

The authors declare no competing interests.